\documentclass{iau}
\usepackage{graphicx}

\title[IAUS 292.~~Molecular Clouds] 
{Molecular Clouds: Internal Properties,\\ Turbulence, Star Formation and Feedback}

\author[J. C. Tan, S. N. Shaske, \& S. Van Loo]   
{Jonathan C. Tan$^1$,
Suzanne N. Shaske$^2$ \and Sven Van Loo$^3$}

\affiliation{$^1$Depts. of Astronomy \& Physics, University of Florida, Gainesville, FL 32611, USA\\
$^2$Dept. of Chemical Engineering, University of Florida, Gainesville, FL 32611, USA\\
$^3$Harvard-Smithsonian Center for Astrophysics, 60 Garden Street, Cambridge, MA 02138, USA}

\pubyear{2012}
\volume{292}  
\pagerange{1--8}
\jname{Molecular Gas, Dust, and Star Formation in Galaxies}
\editors{T. Wong \& J. Ott, eds.}
\begin{document}

\maketitle

\begin{abstract}
All stars are born in molecular clouds, and most in giant molecular
clouds (GMCs), which thus set the star formation activity of galaxies.
We first review their observed properties, including measures of mass
surface density, $\Sigma$, and thus mass, $M$. We discuss cloud
dynamics, concluding most GMCs are gravitationally bound.
Star formation is highly clustered within GMCs, but overall is very
inefficient. We compare properties of star-forming clumps with those
of young stellar clusters (YSCs). The high central
densities of YSCs may result via dynamical evolution of already-formed
stars during and after star cluster formation. We discuss theoretical
models of GMC evolution, especially addressing how turbulence is
maintained, and emphasizing the importance of GMC collisions. We
describe how feedback limits total star formation efficiency,
$\epsilon$, in clumps. A turbulent and clumpy medium allows higher
$\epsilon$, permitting formation of bound clusters even when escape
speeds are less than the ionized gas sound speed.

\end{abstract}

\firstsection 
\section{Observed Properties of Molecular Clouds}\label{S:obs}

Stars form from molecular gas, which, in the Milky Way, is mostly in
GMCs, defined to have $M\geq 10^4\:M_\odot$ (Williams et
al. 2000). Within GMCs are overdense {\it clumps} that may later form
star clusters and {\it cores} that may form single stars or
binaries. GMCs are normally identified by $^{12}$CO(1-0) emission
(e.g. Solomon et al. 1987 [S87]) as discrete features in
$l-b-v$ space. They have a small volume fraction of the Galactic
disk, but a significant mass fraction, $\sim 1/3$, of gas inside
the solar circle (Wolfire et al. 2003).

CO is dissociated by FUV photons, so CO-defined GMCs have a protective
layer of molecular ``dark gas'' (Grenier et al. 2005). For the local
FUV field, dust in two layers on each side of the cloud provide
$A_V\sim 1.4$~mag (van Dishoeck \& Black 1988), i.e. $\Sigma=6.2\times
10^{-3}\:{\rm g\:cm^{-2}} = 30\:M_\odot\:{\rm pc}^{-2}$ or $N_{\rm
  H}=2.7\times 10^{21}\:{\rm cm^{-2}}$ (we adopt $n_{\rm He}/n_{\rm
  H}=0.1$, so mass per H, $\mu_{\rm H} = 2.34\times 10^{-24}\:{\rm
  g}$). Figure~1, a diagram of $\Sigma$ vs. $M$ for molecular clouds
and star clusters, indicates this level. GMCs have larger mean $\Sigma
\sim 0.02\:{\rm g\:cm^{-2}}$ (or $\sim 100\:M_\odot\:{\rm pc}^{-2}$)
(\S\ref{S:GMC}). Nevertheless, significant fractions, $\sim 0.3$, of
total GMC molecular masses are predicted to be in the dark gas phase
(Wolfire et al. 2010). The porous nature of GMCs increases FUV
penetration and a large fraction, perhaps as high as 90\%, of
molecular gas can be considered to be part of Photodissociation
Regions, where the FUV radiation field influences ionization,
heating and chemistry (Hollenbach \& Tielens 1999). The predicted
equilibrium temperatures in the high extinction clumps within GMCs are
$\sim 10$~K, corresponding to sound speeds of $c_s = (kT/\mu)^{1/2} =
0.188 (T/10\:{\rm K})^{1/2}\:{\rm km\:s^{-1}}$.

\subsection{Methods to Measure Mass Surface Density, $\Sigma$, of Molecular Clouds}\label{S:Sigma}

(1) From $^{12}$CO via the ``X-factor'': $X\equiv N_{\rm H_2}/I_{\rm
  CO(1-0)}$. Dame et al. (2001) estimate $X\simeq (1.8\pm0.3) \times
10^{20}\:{\rm cm^{-2}/(K\:km/s)}$ by first finding the ratio of FIR
dust emission to total (HI-derived) column density, $I_{\rm 100\mu
  m}/N_{\rm H}$, around local, high latitude GMCs. This is used
together with $I_{\rm 100\mu m}$ observed at the GMC, to predict the
total $N_{\rm H}$ at its location. The observed contribution from
atomic gas is subtracted and the remainder compared to $I_{\rm
  CO(1-0)}$. Similar results are found using observed $\gamma$-ray
emission (produced by cosmic rays interactions) to estimate total
$N_{\rm H}$ towards GMCs (e.g. Strong \& Mattox 1996; Ackermann
et al. 2011), which are then compared to HI and CO observations of
these locations. Note, these methods allow detection of dark
gas. Leroy et al. (2011) used the FIR emission method to measure $X$
in 5 nearby galaxies, finding it increased by a factor of $\sim 10$ in
the lowest metallicity systems, NGC 6822 \& SMC --- interpreted
as being due to the increasing dominance of dark gas in these
relatively dust-free environments.

(2) From $^{13}$CO(1-0), typically assuming its excitation
temperature, $T_{\rm ex},$ is the same as that derived from
$^{12}$CO(1-0) (assumed to be optically thick). If the $^{13}$CO is
subthermally excited, then $\Sigma$ will be underestimated. Note,
Pineda et al. (2008) find that 60\% of even the $^{12}$CO(1-0)
emission in the Perseus GMC is subthermally excited. Given an adopted
$T_{\rm ex}$, $N_{\rm ^{13}CO}$ can be derived, including corrections
for its optical depth. Then the total $N_{\rm H_2}$ follows, given
an assumed abundance of $^{13}$CO with respect to $\rm H_2$:
e.g. Roman-Duval et al. (2010, RD10) assume $n_{\rm ^{12}CO}/n_{\rm
  ^{13}CO}=45$ (Langer \& Penzias 1990) and $n_{\rm ^{12}CO}/n_{\rm
  H_2}=8\times 10^{-5}$ (Blake et al. 1987),
so $n_{\rm ^{13}CO}/n_{\rm H_2}=1.78\times 10^{-6}$.
$n_{\rm ^{12}CO}/n_{\rm ^{13}CO}$ is expected to vary with
Galactic radius from $\sim40$ at $r=3$~kpc to $\sim 70$ at
$8$~kpc (Milam et al. 2005). $n_{\rm ^{12}CO}/n_{\rm H_2}$
is uncertain and may also show factor of $\sim 2$ variations:
e.g. Lacy et al. (1994) find $n_{\rm ^{12}CO}/n_{\rm H_2}\sim (3\pm1)\times
10^{-4}$. Hernandez \& Tan (2011) adopted $n_{\rm ^{12}CO}/n_{\rm
  ^{13}CO}=54$ and $n_{\rm ^{12}CO}/n_{\rm H_2}=2\times 10^{-4}$ so
$n_{\rm ^{13}CO}/n_{\rm H_2}=3.70\times 10^{-6}$ in their study of two
infrared dark clouds (IRDCs). With these choices and our
preferred value of $\mu_{\rm H}$, the values of $\Sigma$ and $M$ of
RD10 should be scaled by a factor 0.49 (\S\ref{S:GMC} and Fig.~1).

(3) From other molecular line tracers, e.g. $\rm C^{18}O$, $\rm
HCO^+$, $\rm CS$. The rarer isotopologues of CO, being harder to
detect, tend to be observed towards higher $\Sigma$ clumps and cores
within GMCs. A problem in the cold, dense conditions of IRDCs is CO
freeze-out onto dust grains and thus its depletion from the gas
phase. Widespread (extending over several pc) line-of-sight depletion
factors, $f_D$, (i.e. expected gas phase abundance relative to that
observed) of $\sim3$ were seen via $\rm C^{18}O$ compared to $\Sigma$
derived via MIR extinction mapping towards an IRDC by Hernandez et
al. (2011). Larger values of $f_D\sim 30$ were reported by Fontani et
al. (2012) in more localized IRDC clumps via $\rm C^{18}O$(3-2)
compared to $\Sigma$ derived via FIR/submm emission. $\rm HCO^+(1-0)$
is often bright, but has an uncertain and likely variable abundance
with respect to $\rm H_2$. It has a critical density $n_{\rm crit}=1.7
\times 10^5\:{\rm cm^{-3}}$, but line trapping can lower the effective
$n_{\rm crit}$ to $\sim$~few$\times 10^3\:{\rm cm^{-3}}$, similar to
that of optically thin $\rm C^{18}O(1-0)$ (Evans 1999). CS can also be
used to trace dense gas, especially utilizing the multiple higher $J$
transitions observable from the ground (e.g. Plume et al. 1997).

(4) From FIR/mm emission, requiring knowledge of dust
temperature(s), $T_d$, 
and opacity per unit gas mass, $\kappa_\nu$. Assuming optically
thin transfer and black body dust emission, $\Sigma = 4.35\times
10^{-3} ([S_\nu/\Omega] / [{\rm MJy/sr}]) (\kappa_\nu/[0.01{\rm
    cm^2/g}])^{-1}\lambda_{1.2}^3 [{\rm exp}(0.799 T_{d,15}^{-1}
  \lambda_{1.2}^{-1})-1]\:{\rm g\:cm^{-2}}$, where
$\lambda_{1.2}=\lambda/1.2\:{\rm mm}$ and $T_{d,15}=T_d/15\:{\rm
  K}$. A common choice of $\kappa_\nu$ is that due to the moderately
coagulated thin ice mantle dust model of Ossenkopf \& Henning (1994,
OH94), which has an opacity per unit dust mass of $\kappa_{{\rm
    1.2mm},d}=1.056\:{\rm cm^2\:g^{-1}}$. A
gas-to-refractory-component-dust-mass ratio of 141 is estimated by
Draine (2011), while 156 was assumed in the OH94 coagulation model, so
$\kappa_{\rm 1.2mm}=6.77\times 10^{-3}\:{\rm cm^2\:g^{-1}}$. This
method probes all mass along the line of sight, and thus is also
sensitive to dark molecular and atomic gas. In the Galactic plane, the
method gains utility for high $\Sigma$ clumps and cores within GMCs,
which dominate the total line-of-sight column density. However,
accounting for the surrounding diffuse emission is still needed
(Battersby et al. 2011).

(5) From NIR extinction mapping, using background stars (e.g. Lada et
al. 2007; Goodman et al. 2009), probing total line-of-sight $\Sigma$
(up to $A_V \sim 25$ mag). This method depends on dust opacity in the NIR and
the gas-to-dust mass ratio, but not $T_d$. It has first, and most
accurately, been applied to nearby GMCs that sit at relatively high
Galactic latitudes. Kainulainen et al. (2011) used this method to
study clouds in the Galactic plane, relying on statistical binning of
sources in resolution elements of $\sim 30'' \times 30''$.

(6) From MIR extinction mapping, using diffuse Galactic background
emission from hot dust. {\it Spitzer} $8\:{\rm \mu m}$ images resolve
to $2''$ and probe high-$\Sigma$ ($\sim 0.5\:{\rm g\:cm^{-2}}$;
$A_V\sim 100$~mag) IRDCs (e.g. Butler \& Tan 2009 [BT09]; 2012 [BT12];
Peretto \& Fuller 2009). The method depends on $\kappa_{\rm 8\mu m}$
(BT09 use $7.5\:{\rm cm^{2}\:g^{-1}}$ [OH94]), but not
$T_d$. Allowance is made for foreground emission, best measured by
finding ``saturated'' intensities towards independent, optically thick
cores (BT12). Only differences in $\Sigma$ relative to local
surroundings are probed, so the method is insensitive to low-$\Sigma$
environs of IRDCs. This limitation is addressed by combining NIR \&
MIR extinction maps (Kainulainen \& Tan 2012 [KT12]).

Once $\Sigma$ is measured, cloud mass, $M$, is found from projected
angular area and an estimate of distance, typically kinematic for
Galactic clouds. In the inner Galaxy, the near/far degeneracy may be
broken by looking for HI self-absorption. For IRDCs, the near
kinematic distance is often assumed. Kinematic distances are uncertain
due to non-circular cloud motions caused by their $\sim 5-10\:{\rm
  km\:s^{-1}}$ disk-plane velocity dispersion or streaming due to
spiral arms or bars. Parallax distances to masers associated with star
formation are available for some sources (see, e.g. Reid et al. 2009;
Foster et al. 2012).

\subsection{Virial Equilibrium of Molecular Clouds}\label{S:virial}

The 1D internal velocity dispersion, $\sigma$, of clouds is measured
from their line widths. This allows assessment of cloud kinetic
energy, ${\cal T} = (3/2)M\sigma^2$, and comparison to gravitational
energy, $W = -(3/5)a GM^2/R$, where $a$ is a factor, $\sim{\cal O}(1)$,
that accounts for nonspherical and nonuniform density distributions
(Bertoldi \& McKee 1992 [BM92]). BM92 find the effects of
nonsphericity are quite small, so we neglect them here. For
a power-law density distribution $\rho\propto r^{-k_\rho}$, $a =
(1-k_\rho/3)/(1-2k_\rho/5) = 10/9, 5/4, 5/3$ for $k_\rho=1,3/2,2$,
respectively. BM92 define the virial parameter as $\alpha_{\rm
  vir}\equiv 5\sigma^2R/(GM) = 2 a {\cal T}/|W|$. A cloud in virial
equilibrium satisfies $2({\cal T}-{\cal T}_0) + {\cal M} + W = 0$,
where ${\cal T}_0= (3/2) P_0 V$ ($P_0$ is the surface pressure around
the cloud volume, $V$) and ${\cal M} = [1/(8\pi)]\int (B^2-B_0^2) dV$
($B_0$ is the magnetic field far from the cloud and the integral
extends beyond the cloud surface over regions where the B-field
suffers distortions).

Molecular clouds are magnetized (Crutcher 2012), with observed values
being consistent with a uniform distribution of total field strengths
from 0 to a maximum, $B_{\rm TOT,0} = 10\:\mu G$ for $n_{\rm
  H}<300\:{\rm cm^{-3}}$, and $B_{\rm TOT,0} = 10 (n_{\rm H}/300\:{\rm
  cm^{-3}})^{0.65}$ for $n_{\rm H}>300\:{\rm cm^{-3}}$. However, in
general we do not have accurate estimates of total field strengths in
and around a given GMC or clump, making it difficult to assess its
contribution to virial balance.

For a cloud satisfying the equilibrium virial equation, $\alpha_{\rm
  vir} = a(1+(2{\cal T}_0 - {\cal M})/|W|) = a
(1-P_0/\bar{P})^{-1}(1-{\cal M}/|W|)$, where $\bar{P}\equiv
(3/2)\bar{\rho}\sigma^2$ is the volume-averaged cloud pressure. For
negligible surface pressure ($P_0\rightarrow 0$) and magnetic support
(${\cal M}\rightarrow 0$), $\alpha_{\rm vir} = a$, i.e. ${\cal
  T}=|W|/2$. Then, a gravitationally bound cloud
has $\alpha_{\rm vir}<2$. If a cloud has $P_0=\bar{P}/2$, $\alpha_{\rm
  vir}$ is raised by a factor of 2. If ${\cal M}=|W|/2$, $\alpha_{\rm
  vir}$ is lowered by a factor of 2.


\subsection{Giant Molecular Clouds}\label{S:GMC}

For $^{12}$CO-defined GMCs, S87 measured $\bar{\Sigma} \sim
170\:M_\odot\:{\rm pc^{-2}}$. However, Heyer et al. (2009, [H09])
measured $\bar{\Sigma} \sim 42\:M_\odot\:{\rm pc^{-2}}$ for these same
structures using $^{13}$CO. As discussed by H09, the true $\Sigma$ is
likely to be lower than the S87 estimate due to incomplete sampling of
the original $^{12}$CO maps, but the $^{13}$CO value may be
underestimated because of subthermal excitation. We indicate the
properties of Galactic $^{12}$CO-defined GMCs in Fig.~1 with
$\Sigma\sim 100\:M_\odot\:{\rm pc^{-2}}$ and a mass range from
$10^4\:M_\odot$ to $\sim 6\times 10^6\:M_\odot$ (Williams \& McKee
1997). Several outer-Galaxy GMCs and their clumps were studied by
Heyer et al. (2001), who found they had $\bar{\Sigma} \sim
10\:M_\odot\:{\rm pc^{-2}}$. Bolatto et al. (2008) studied GMCs in
nearby galaxies, finding they generally had properties similar to
Galactic GMCs. However, GMCs may differ in more extreme environments:
e.g. 14 GMCs observed in the dwarf starburst Henize 2-10 have
$\bar{\Sigma}\simeq 400\:M_\odot\:{\rm pc^{-2}}$ (Santangelo et
al. 2009 [S09]).

The $^{13}$CO-defined Galactic GMCs and clumps of RD10 (with $\Sigma$
\& $M$ scaled by 0.49, [\S\ref{S:Sigma}]) are also shown in
Fig.~1. For clouds with $M>10^4\:M_\odot$, $\bar{\Sigma} = 124
M_\odot\:{\rm pc^{-2}}$, including dark gas.  For this sample,
considering $\alpha_{\rm vir}$ reported by RD10 (increased by factors
2.02 to account for the mass scaling [\S\ref{S:Sigma}] and 1.28 to
match the BM92 definition, for total of 2.60; note, we neglect
contribution of dark gas to $\alpha_{\rm vir}$) we find
$\bar{\alpha}_{\rm vir}=1.09$. Also, 89\% of these 249 GMCs (with 95\%
of the mass) have $\alpha_{\rm vir}<2$, i.e. are gravitationally
bound. Similar results are obtained for $\rm ^{12}CO$-defined GMCs, if
one allows for a factor $\sim$2 increase in H09 masses (above) and for
the extragalactic GMCs studied by Bolatto et al. (2008).

GMCs that are gravitationally bound and near virial equilibrium explain
H09's result of a modified line-width vs. size relation: $\sigma
 \propto (\Sigma R)^{0.5}$, which follows if $M\simeq M_{\rm vir}\simeq
5\sigma^2R/G$ and $\Sigma = M/(\pi R^2)$. Note, this may also explain
the higher normalization of the relation seen in high-$\Sigma$ clumps
(e.g. Plume et al. 1997), if they are also self-gravitating. 

Thus we conclude that GMCs are gravitationally bound and have
configurations that are consistent with being in approximate
gravitational virial equilibrium ($\alpha_{\rm vir}\simeq 1$).



\begin{figure}[b]
\vspace*{-1.35 cm}
\begin{center}
 \includegraphics[width=5.1in]{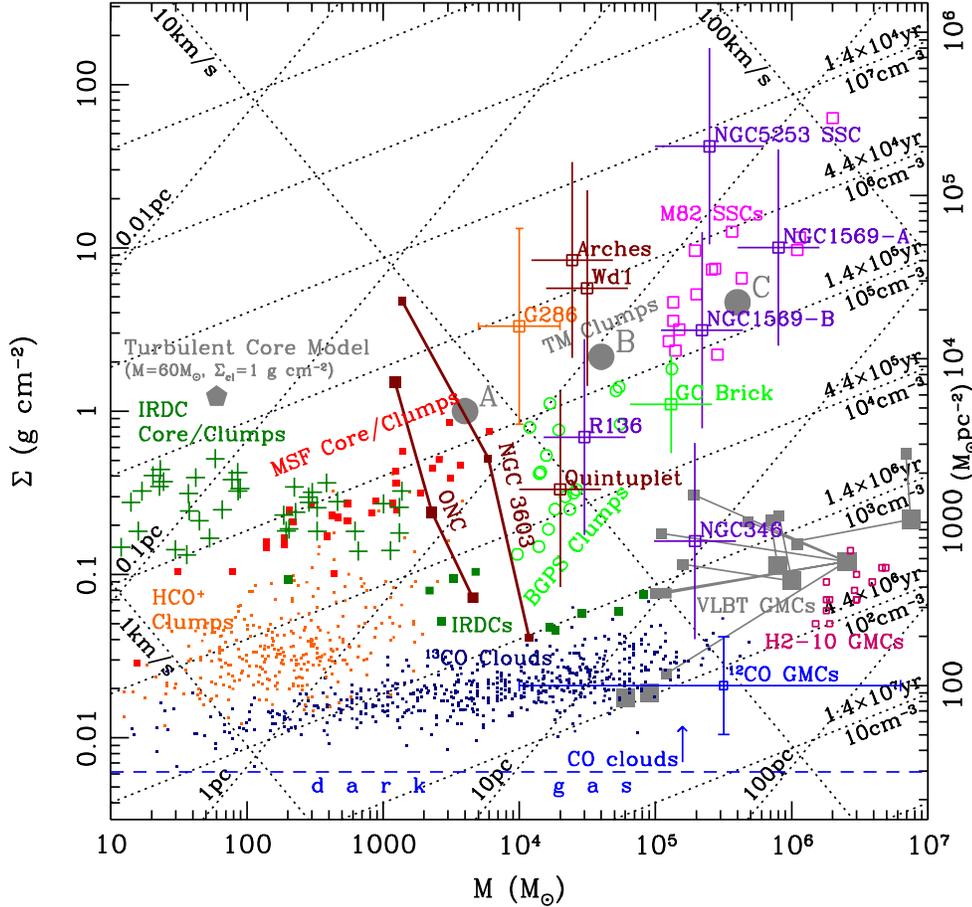} 
\vspace*{-0.45 cm}
 \caption{ 
$\Sigma$-$M$ Diagram of Molecular Clouds \& Young Star Clusters. 
Mass surface density, $\Sigma\equiv M/(\pi R^2)$, is plotted vs.
mass, $M$. Dotted lines of constant radius, $R$, H number
density, $n_{\rm H}$ (or free-fall time, $t_{\rm ff}=
[3\pi/(32G\rho)]^{1/2}$), and escape speed, $v_{\rm esc} = (10/\alpha_{\rm vir})^{1/2}\sigma$, are
shown. The minimum $\Sigma$ for local CO-emitting
clouds due to FUV radiation is $\sim 30\:M_\odot\:{\rm pc^{-2}}$
(dashed line). Below this molecular ``dark gas'' can exist. The
following objects are plotted (see text for more details). Typical
$\rm ^{12}CO$-defined GMCs have $\Sigma\sim 100\:M_\odot\:{\rm
  pc^{-2}}$, although denser examples have been found in Henize 2-10
(S09). The $\rm ^{13}CO$-defined clouds of RD10 are indicated, along
with $\rm HCO^+$ clumps of B11, including G286 (B10). 10 IRDCs (KT12)
and their internal core/clumps (BT12) are shown, the latter overlapping with Massive Star-Forming core/clumps [M02]. Along with
G286, the BGPS clumps (G12) and the Galactic Center Brick (L12) are
some of the most massive, high-$\Sigma$ gas clumps known. Clumps may
give rise to young star clusters (see text for refs.), like the ONC
and NGC 3603 (radial structure shown from core to half-mass,
$R_{1/2}$, to outer radius), or even more massive examples,
e.g. Westerlund 1, Arches, Quintuplet (shown at $R_{1/2}$). Example
clusters in the LMC (R136) \& SMC (NGC 346) show a wide range of
$\Sigma$.
More massive, Super Star Clusters are found in some dwarf
irregular galaxies, NGC 1569 \& 5253, and starburst galaxy M82. Grey
squares indicate the GMCs simulated by VLBT12 from beginning (large)
to end (small) of a 10~Myr period, including merger and fragmentation
of clouds. Grey circles, A, B, C, show TM01 feedback model clumps, yielding star formation efficiencies of
$\epsilon=0.33, 0.48, 0.50$, respectively. Grey pentagon shows initial condition of a fiducial
MSF turbulent core of MT03.
}
\label{fig1}
\end{center}
\end{figure}

\subsection{Clumps and Cores}


Barnes et al. (2011 [B11]) derived the properties of a complete census
of $\sim$300 ${\rm HCO^+}$ clumps in a $20^\circ\times 6^\circ$ region
of the Galactic plane (Fig.~1). While the absolute properties of
individual sources are uncertain due to the assumed ${\rm HCO^+}$
abundance and $T_{\rm ex}$, relative properties should be more
accurate. The most extreme (in $M$ \& $\Sigma$) clump in this sample
is G286 (Barnes et al. 2010 [B10]), which also contains an embedded,
forming cluster.

Ginsburg et al. (2012 [G12]) presented properties of a census of
massive ($\gtrsim 10^4\:M_\odot$) clumps found by 1.1~mm emission
in the Bolocam Galactic Plane Survey (BGPS). In Fig.~1 we have scaled
their (free-free-subtracted, $T_d=20$~K) masses by a factor of 1.47 to
account for our preferred choice of $\kappa_{\rm 1.1mm}=7.78\times
10^{3}\:{\rm cm^{2}\:g^{-1}}$ (OH94). Longmore et al. (2012 [L12])
reported a comparable, cold, $1.3\times 10^5\:M_\odot$, 2.8~pc-radius
Galactic center clump, dubbed the ``Brick''. It shows little evidence
of star formation activity.

Cold, high-$\Sigma$ regions of GMCs reveal themselves as IRDCs and 10
such clouds are shown in Fig.~1, with masses derived from the combined
NIR+MIR extinction mapping method of KT12. We also show 42 core/clumps
selected from these IRDCs by BT12, (properties derived for
best-fitting singular polytropic sphere model). The structure is
resolved interior to these radii down to $2''$ ($\simeq 0.04$~pc at
4~kpc). The mean value of $k_\rho = 1.1\pm 0.25$. If account is taken
of the surrounding clump envelope, then $k_\rho = 1.6\pm 0.28$. Some
of these objects, as well as others reported by Bontemps et
al. (2010), appear to be consistent in their properties with massive
starless cores invoked in the Turbulent Core Model of massive
star formation (MSF) (McKee \& Tan 2003 [MT03]).

Mueller et al. (2002 [M02]) studied dense cores and clumps, selected by
water maser activity to harbor MSF. For 31 sources, they derived the
density and temperature structure by 1D radiative transfer modeling of
the IR/submm emission, finding $k_\rho \simeq 1.8\pm0.4$ (however,
these results may be affected by the presence of bipolar outflow
cavities, Zhang \& Tan 2011). 
For consistency, in Fig. 1 we have multiplied the M02 values of
$\Sigma$ and $M$ by a factor of 1.56, and we show their properties at
a scale where $n_{\rm H}\sim 3\times 10^4\:{\rm cm^{-3}}$.


Most lower $M$ and $\Sigma$ ${\rm ^{13}CO}$ and $\rm HCO^+$ clumps
appear pressure confined (e.g. BM92, B11), i.e. $\alpha_{\rm
  vir}\gg 1$ and their internal pressure is influenced more by the
ambient pressure rather than by self-gravitating weight. Note,
the clumps may still satisfy the equilibrium virial
equation. Higher-$\Sigma$ clumps appear to be nearer gravitational virial
equilibrium (i.e. internal pressure set by self-gravity to
be $\sim G\Sigma^2$). This has been found in IRDC filaments  
(Hernandez et al. 2012), clumps (KT12) and cores 
(Tan et al., in prep.). In 
G286, B10 report a mass infall rate $3.4\times
10^{-2}\:M_\odot\:{\rm yr^{-1}}$, highest in the B11 sample, but
still $\sim$10 times less than the rate expected under global
free-fall collapse.

\subsection{Star Formation: Embedded and Young Star Clusters (YSCs)}

The star formation efficiency per $t_{\rm ff}$ of GMCs and clumps is
puny, $\epsilon_{\rm ff}\sim 2\%$ (Zuckerman \& Evans 1974; Krumholz
\& Tan 2007). It is unclear whether this is due mostly to support by
turbulence (Krumholz \& McKee 2005) or magnetic fields, although
evidence for a threshold $\Sigma\sim 120\:M_\odot\:{\rm pc^{-2}}$
($A_V\sim 8$~mag) for star formation (Lada et al. 2010) may indicate
the importance of the latter (McKee 1989). Star formation is
concentrated in clumps in GMCs, where total efficiencies
$\epsilon\gtrsim 0.3$ can be achieved, as evidenced by bound YSCs. A
ratio $\epsilon_{\rm ff}/\epsilon \ll 1$ implies cluster formation
times, $t_{\rm form} \gg t_{\rm ff}$ (Tan et al. 2006).

Embedded star clusters provide a snapshot of the star cluster
formation process. YSCs may also retain memory of the physical
conditions of the gas clumps from which they formed, although modified
by mass segregation, cluster core collapse, expansion due to gas
massloss, and tidal stripping (see Portegies-Zwart et al. 2010
[PZMG10]). In Fig.~1 we show the Orion Nebula Cluster (ONC), with
density structure from Hillenbrand \& Hartmann (1998) with $M$ and
$\Sigma$ estimated at the core, half-mass, $R_{1/2}$, and total
radii. These masses are virial mass estimates, and so receive
contribution from both stars and gas. Similarly, we show NGC 3603 ---
a significantly more massive cluster (Harayama et al. 2008). The most
massive known Galactic YSC is Westerlund 1, with dynamical mass $\sim
6.3^{+5.3}_{-3.7}\times 10^4\:M_\odot$ and $R_{1/2}=$0.86~pc (Mengel
\& Tacconi-Garman 2007). Harfst et al. (2010) modeled the expected
initial conditions of the Arches cluster, near the Galactic Center, to
match its observed stellar density distribution 2.5~Myr later. Their
preferred best-fit model has an initial mass of $(4.9\pm0.8)\times
10^4\:M_\odot$ (IMF uncertainties introduce larger errors) and a
virial radius of 0.70~pc (i.e. $R_{1/2}\simeq 0.44$~pc). We show this
half-mass condition in Fig.~1. The Quintuplet cluster currently has
about 6,000~$M_\odot$ inside a radius of 0.5~pc (Hu\ss{}mann et
al. 2012), but being $\sim$4~Myr old, it is more difficult to
reconstruct its initial conditions: we adopt $M=4\times 10^4\:M_\odot$, $R_{1/2}=$2~pc (Figer et al. 2006; PZMG10).

In the 30 Doradus region of the LMC, R136 has $M\simeq 6\times
10^4\:M_\odot$ (Andersen et al. 2009) and $R_{1/2}=$1.7~pc (Hunter et
al. 1995), similar to the dynamical mass estimate of H\'enault-Brunet
et al. (2012). NGC 346 is a much lower density star cluster (or
association) in the SMC, with $M=3.9\times 10^5\:M_\odot$ and
$R_{1/2}=9$~pc. More massive, ``Super Star Clusters'' (SSCs) have been
found in, e.g., the dwarf irregular galaxies NGC~1569 (cluster A with
$M=1.6\times 10^6\:M_\odot$, $R_{1/2}=2.3$~pc, Anders et al. 2004,
PZMG10; and B with $M=4.4\times 10^4\:M_\odot$, $R_{1/2}=2.2$~pc,
Larsen et al. 2008) and NGC~5253 ($M\sim 5\times 10^5\:M_\odot$,
$R_{1/2}\sim 0.6$~pc, Turner \& Beck 2004) and in M82 (McCrady \&
Graham 2007) (Fig.~1).

A comparison of YSCs and gas clumps in the $\Sigma-M$ diagram (Fig.~1)
suggests relatively massive starless (or at least gas-dominated)
clumps do form, albeit rarely, that may be progenitors of the most
massive Galactic clusters. These can have $\bar{n}_{\rm H}\sim 10^5\:{\rm
  cm^{-3}}$ and $t_{\rm ff}\sim 10^5\:{\rm yr}$. Star formation
appears inhibited during assembly of these massive clumps. We suspect
this is due to support by magnetic fields rather than turbulence,
which would quickly decay (\S\ref{S:theory}). It is not yet
clear if $\sim 10^6\:M_\odot$ starless clumps can form to be
progenitors of SSCs. Central core densities of many YSCs exceed
what is seen in gas clumps and may be the result of dynamical
evolution of the cluster both during the star cluster formation
process (especially if $t_{\rm form}\gg t_{\rm ff}$) and after star
formation has ceased.


\vspace{-0.48cm}
\section{Theoretical Models of Molecular Clouds}\label{S:theory}

GMCs and clumps exhibit density structures on a wide range of
scales, all the way down to those of forming stars. They contain
motions, initially imprinted from the surroundings or due
to self-gravity, and later due to feedback processes, that are many
times the sound speed, leading to strong shocks. They are
magnetized. They have poorly-defined boundaries, merging
continuously into the surrounding galactic environment.
These features
make theoretical and numerical modeling of molecular clouds 
a challenging problem.

\subsection{Turbulence and GMC Evolution}

\begin{figure}[b]
\vspace*{-0.55 cm}
\begin{center}
\includegraphics[width=1.92in]{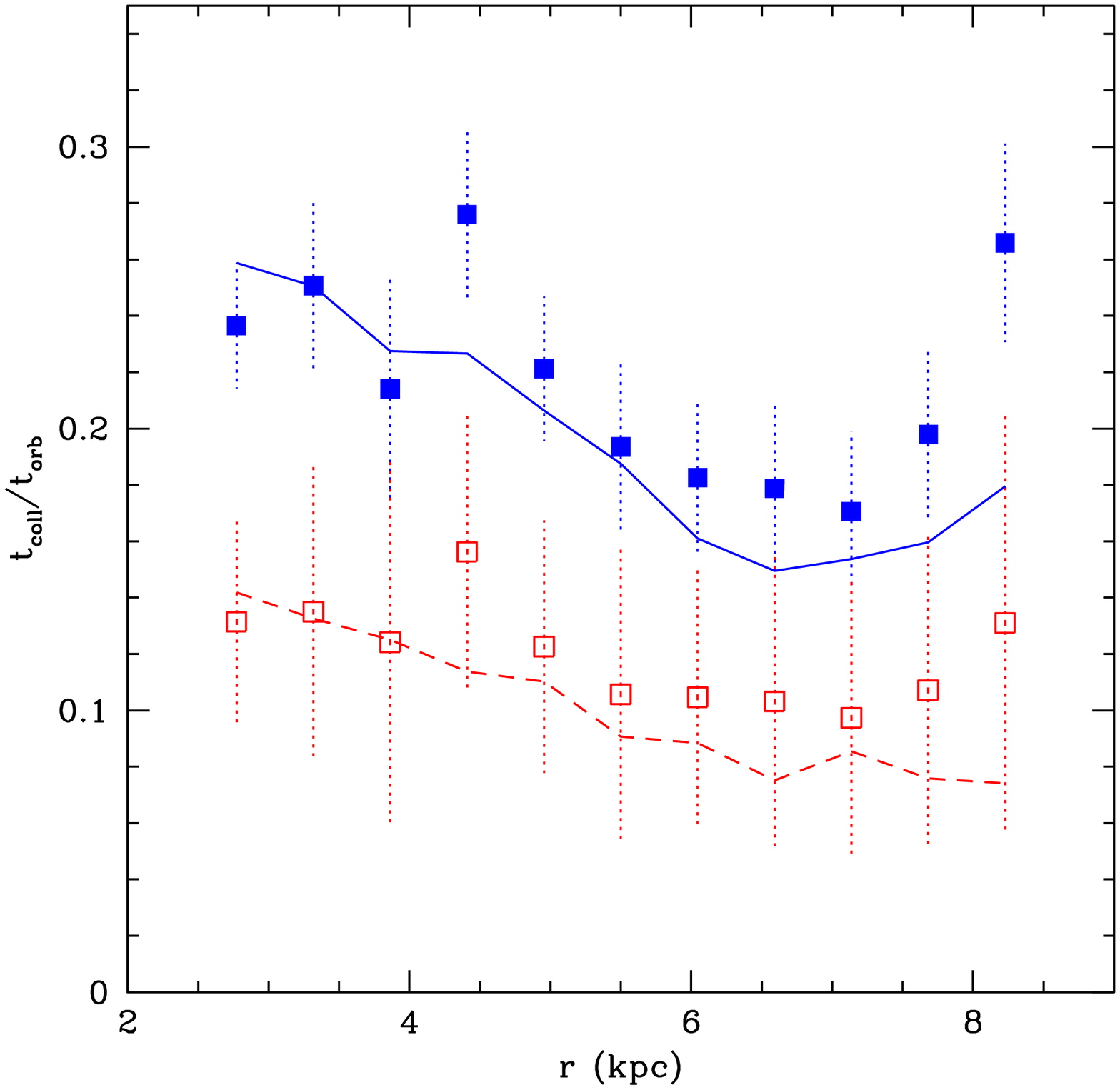} 
\includegraphics[width=3.35in]{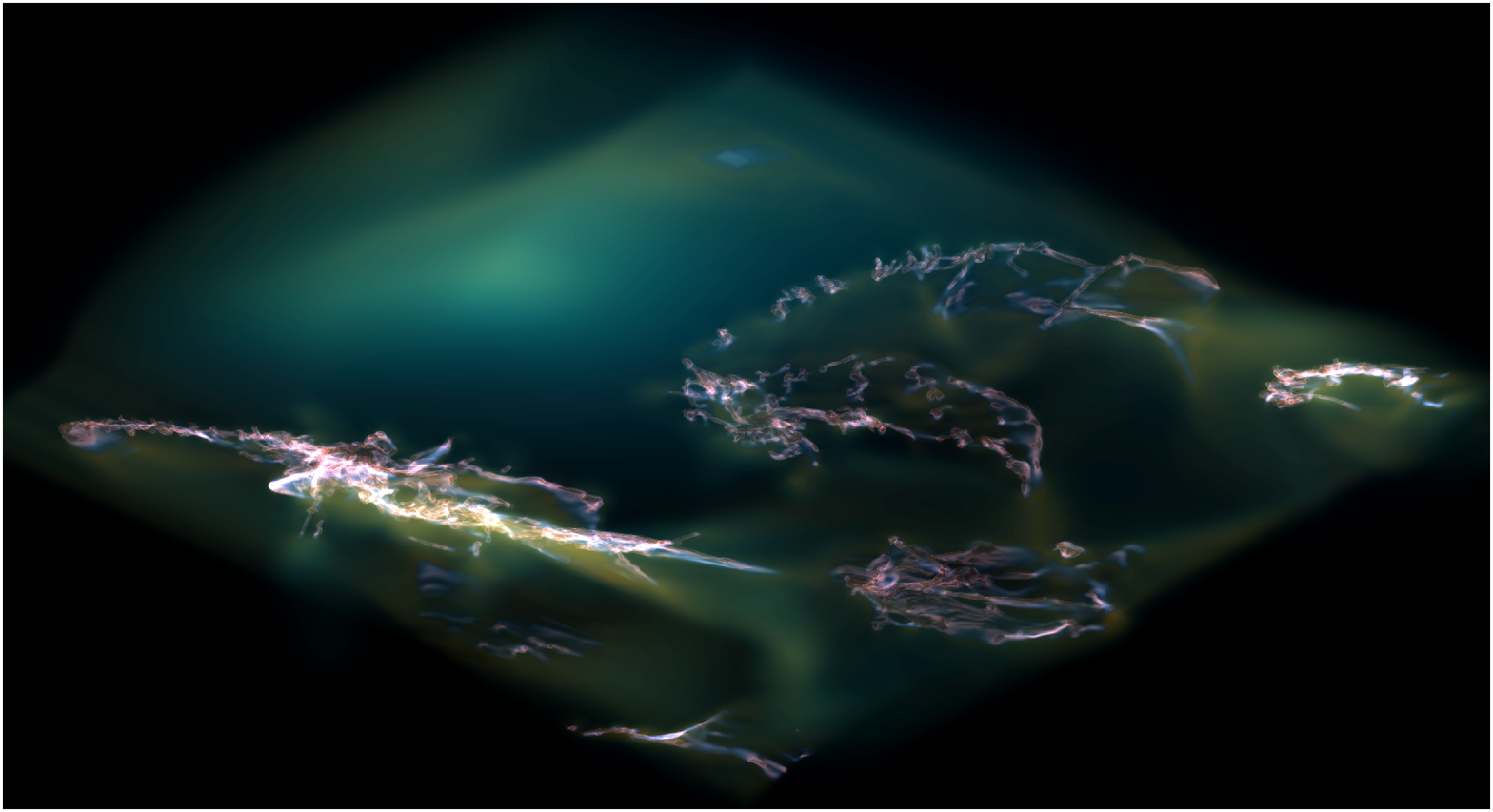} 
\vspace*{-0.55 cm}
\caption{
(a) 
Left: GMC-GMC collision time, $t_{\rm coll}$, relative to local
orbital time, $t_{\rm orb}$ vs. $r$ in the TT09 simulation for all
(solid line) and $M>10^6\:M_\odot$ (dashed line) GMCs. Solid/open
squares show the analytic estimate (eq.~\ref{eq:tcoll}) for these
respective populations (vertical lines show dispersion of the
averages sampled 50 times at 1~Myr intervals. (b) Right: Volume
density rendering ($n_{\rm H}\sim 100\:{\rm cm^{-3}}$ - blue;
$10^5\:{\rm cm^{-3}}$ - red) of a 0.5~pc-resolution simulation of
1~kpc patch of a galactic disk (VLBT12), extracted from the TT09
simulation and evolved for 10~Myr.}
\label{fig2}
\end{center}
\end{figure}

GMCs velocity dispersions are $\gg c_s$.
Supersonic turbulence decays in $\sim 1$ dynamical time, $t_{\rm
  dyn}=R/\sigma$ (Stone et al. 1998; Mac Low et al. 1998) (for
$\alpha_{\rm vir}\simeq 1$, $t_{\rm ff} \simeq 0.5 t_{\rm
  dyn}$). Maintenance of turbulence is a constraint on models of GMC
formation and evolution. McKee \& Ostriker (2007 [MO07]) reviewed two
conceptual frameworks: (1) GMCs are dynamic, transient and largely
unbound, with turbulence driven by large-scale colliding
atomic flows that form the clouds (e.g. Heitsch et al. 2005;
V\'azquez-Semadeni et al. 2011). These models do not
explain why most GMCs are bound with $\alpha_{\rm vir}\sim 1$
(\S\ref{S:GMC}). Fast flows of HI around GMCs needed to form them
quickly in $\sim t_{\rm dyn}$ have not been observed. Nor is it clear
if such models are globally consistent with the relatively high mass
fractions of gas in GMCs, at least inside the solar circle. (2) GMCs
form by large-scale gravitational instabilities and are thus
gravitationally bound with $\alpha_{\rm vir}\sim 1$. Turbulence is
maintained by contraction and then, later, star formation feedback
(e.g. Goldbaum et al. 2011).

Here we present a third paradigm: GMC evolution under frequent cloud
collisions (Tan 2000; Tasker \& Tan 2009 [TT09]; Tan \& Shaske,
in prep.). In a galactic disk with Toomre $Q\sim 1$ and significant
fraction, $f_{\rm GMC}\sim 1/2$, of ISM gas mass in gravitationally
bound clouds, (centers of which appear as GMCs), clouds suffer
collisions on timescale,
\begin{equation}
t_{\rm coll}\simeq [2 b^\prime(\Omega - d v_{\rm circ}/dr){\cal N}_A r_t^2 f_G]^{-1}, 
\label{eq:tcoll}
\end{equation}
where $b^\prime \equiv b/r_t\simeq 1.6$ is typical impact parameter of
collisions (Gammie et al. 1991 [G91]), $\Omega=v_{\rm circ}/r =
2\pi/t_{\rm orb}$, $v_{\rm circ}$ is orbital speed at galactocentric
radius $r$, ${\cal N}_A$ is number of GMCs per unit area,
$r_t=(1-\beta)^{-2/3}(2M/M_{\rm gal})^{1/3}r$ is GMC tidal radius of
GMC at location of given $\beta \equiv d {\rm ln} v_{\rm circ}/ d {\rm
  ln} r$ and interior galactic mass $M_{\rm gal}$, and $f_G\sim 0.5$
is fraction of strong gravitational encounters that lead to
collision. For ${\cal N}_A\simeq f_{\rm GMC} \Sigma_{\rm gas}/\bar{M}$
and $\sigma_{\rm gas}\simeq (G\bar{M}\kappa)^{1/3}(1.0-1.7\beta)$
(G91), we have $t_{\rm coll}\simeq 0.15 Q t_{\rm orb}/(b^\prime f_{\rm
  GMC} f_G [1-0.7\beta])\rightarrow 0.19 f_{\rm GMC}^{-1} t_{\rm orb}$, where the last evaluation is for constant
$v_{\rm circ}$, i.e. $\beta=0$. TT09 found approximate agreement with
this result in their simulation of GMCs in a galactic disk. In Fig.~2a
we evaluate eq.~(\ref{eq:tcoll}) for each TT09 GMC from
$t=225-275$~Myr, and compare the predicted mean $f_{\rm coll}\equiv
t_{\rm coll}/t_{\rm orb}$ with that measured in the simulation. Good
agreement is found for both the entire GMC sample and those with
$M>10^6\:M_\odot$.

We balance the momentum injection rate due to cloud collisions,
$\dot{p}_{\rm CC}$, with the dissipation rate within the clouds,
$\dot{p}_{\rm diss}$, to predict typical GMC density and
$\Sigma$. We have $\dot{p}_{\rm CC}=2 \bar{M} v_{\rm rel}/t_{\rm coll}$,
where the relative velocity $v_{\rm rel}=a v_{s,r_t}=ar_t\Omega$, for which
$v_{s,r_t}$ is the shear velocity at $b=r_t$ and
$a\simeq 1.9$ is expected for typical $b=1.6r_t$ added in quadrature
with $\sigma_{\rm gas}\simeq v_{s,r_t}$. Thus, $\dot{p}_{\rm
  CC}=(a/[\pi f_{\rm coll}])(2G)^{1/3}(\bar{M} v_{\rm circ}/r)^{4/3}$. For turbulent dissipation
we have $\dot{p}_{\rm diss}\equiv -p/(\alpha_{\rm diss}t_{\rm
  ff})=-\sqrt{3}\bar{M}\sigma/(\alpha_{\rm diss}t_{\rm ff}) = -2.19
\alpha_{\rm vir}^{1/2}G\bar{M}\bar{\Sigma}/\alpha_{\rm diss}$. The turbulent energy
dissipation rate has been measured in numerical simulations to be
$\epsilon_{\rm diss} = (1/2)(\dot{E}/{E})[l_0/(\sqrt{3}\sigma)]\simeq
0.6$ (see review by MO07), so that $\alpha_{\rm diss}\simeq 3.85$
(in a cloud with $\alpha_{\rm vir}\simeq 1$ and $l_0=2R$). Thus the
equilibrium value of $\bar{\Sigma}$ for GMCs is
\begin{equation}
\bar{\Sigma}_{\rm eq} = 297 \alpha_{\rm vir}^{-1/2} \frac{\alpha_{\rm diss}}{4} \frac{a}{2}\frac{0.2}{f_{\rm coll}} \left(\frac{\bar{M}}{10^6\:M_\odot}\right)^{1/3} \left(\frac{v_{\rm circ}}{200\:{\rm km\:s^{-1}}} \frac{6\:{\rm kpc}}{r}  \right)^{4/3} M_\odot\:{\rm pc^{-2}}.
\label{eq:Sigma}
\end{equation}

This fiducial value is similar to $\bar{\Sigma}\simeq
300\:M_\odot\:{\rm pc}^{-2}$ of the TT09 GMCs,
suggesting momentum injection from collisions is maintaining their
turbulence and setting their density. Several GMCs extracted from the
TT09 simulation have been evolved to higher, 0.5~pc, resolution by Van
Loo et al. (2012 [VLBT12]) (Fig.~2b). Dense clumps and filaments form,
some reminiscent of IRDCs such as the $\sim
100$~pc-long ``Nessie'' filament (Jackson et al. 2010), but overall
$\bar{\Sigma}$s of the GMCs remain quite constant (Fig.~1), with
clouds supported by a turbulent energy cascade from larger scales,
including via collisions.

Evaluating eq.~(\ref{eq:Sigma}) for Galactic conditions, where $f_{\rm
  GMC}$, $f_{\rm coll}^{-1}$ and $\bar{M}$ are somewhat smaller than
in the TT09 simulation, then we obtain $\Sigma_{\rm eq}$ closer to the
$\sim 100\:M_\odot\:{\rm pc^{-2}}$ of observed GMCs. In addition to helping to
maintaining turbulence, relatively frequent GMC collisions produce a
wide spread in rotation directions (Dobbs 2008; TT09), split roughly
evenly between pro and retrograde with respect to galactic rotation,
as is seen in Galactic (Koda et al. 2006; Imara \& Blitz 2011) and M33
(Imara et al. 2011) GMCs.

\vspace{-0.1cm}
\subsection{Star Cluster Formation and Feedback}

How do so many O stars form together in the small volume of a massive
YSC in the face of their mutual feedback of winds, ionizing
photons and radiation pressure?  
Tan \& McKee (2001 [TM01]; 2004), modeled feedback on 
clumps of given $M$ \& $\Sigma$, idealized as a population of cores
in an intercore medium. To approximate turbulent motions, cores
orbited freely in the clump potential. Stars were formed at
constant rate, $\epsilon_{\rm ff}=0.02$,
with a near-Salpeter IMF, and placed at clump center. Wind,
ionization and radiation pressure feedback were derived from
Starburst99 models.
The spherical average extent of the HII region was calculated
balancing ionizing flux with recombinations, dust absorption and core
shadowing. Cores exposed to ionizing flux were imploded with the
magnetized radiation driven implosion models of Bertoldi \& McKee
(1990). Core dynamics included the photoevaporative rocket effect, as
well as radiation pressure. Advection of neutral cores into the HII
region and subsequent disruption maintained high ionized gas
densities, thus confining the HII region. A momentum conserving
stellar wind bubble swept-out the HII region center, its
boundary set by balancing wind ram pressure with HII
region thermal pressure. Star formation was stopped once the HII region neared
the edge of the clump or after 3~Myr, at the onset of
supernovae. Three example clumps were considered, A, B, C (Fig.~1),
with $M=4\times 10^3, 4\times 10^4, 4\times 10^5\:M_\odot$ and
$\Sigma=1.0,2.2,4.6\:{\rm g\:cm^{-2}}$, resulting in $\epsilon=0.33,
0.48, 0.50$, respectively. Feedback confinement and $\epsilon$ high
enough to form a bound cluster occurred even when $v_{\rm
  esc}\lesssim 10\:{\rm km\:s^{-1}}$, the ionized gas sound speed.

Fall et al. (2010) presented a simpler analytic model that considers
energy and momentum driven feedback on a clump with a smooth radial
density profile. Numerical simulations of feedback interaction with
clumpy and/or turbulent gas have been presented for: ionization by,
e.g., Henney et al. (2009), Gritschneder et al. (2010), and Dale et
al. (2012); for wind/supernova shocks by, e.g., Pittard et al.
(2010); and for radiation pressure by, e.g., Krumholz \& Thompson
(2012). However, the challenge remains of combining and fully
resolving all the processes thought important for star cluster
formation.

\acknowledgements We thank P. Barnes, M. Butler, P. Caselli,
S. Chatterjee, P. Crowther, F. Fontani, A. Hernandez, J. Kainulainen,
S. Kong, B. Ma, C. McKee, L. Smith, E. Tasker, M. Wolfire, B. Wu \& Y. Zhang
for helpful discussions, and S. Skillman for the Enzo-yt visualization
of Fig.~2b. JCT acknowledges support from grants NASA ATP09-0094, NASA
ADAP10-0110 and NSF-CAREER AST-0645412 and REU Supplement (to SNS).

\end{document}